\documentclass{article}
\usepackage{spconf,amsmath,graphicx,hyperref}
\usepackage{booktabs}
\usepackage{siunitx}
\usepackage{amsfonts}
\makeatletter

\let\endOLDthebibliography\endthebibliography
\renewcommand\endthebibliography{%
  \endOLDthebibliography
}

\makeatletter

\newcommand{\inlineSubsection}[2]{
  \refstepcounter{subsection} 
  \noindent\textbf{\thesubsection\ #1}\label{#2}
}

\newcommand{\inlineSubSubsection}[2]{%
  \refstepcounter{subsubsection}%
  \textit{\thesubsubsection\ #1}\label{#2}
}
\usepackage{microtype}
\title{Diffusion Timbre Transfer via Mutual Information Guided Inpainting}
\name{Ching Ho Lee$^{1}$ \qquad Javier Nistal$^{2}$ \qquad Stefan Lattner$^{2}$ \quad Marco Pasini$^{1,2}$ \quad George Fazekas$^{1}$}
  
  \address{$^{1}$ Queen Mary University of London \\
      $^{2}$ Sony Computer Science Laboratories, Paris, France}
\begin{document}
\ninept
\maketitle
\begin{abstract}
 We study timbre transfer as an inference-time editing problem for music audio. Starting from a strong pre-trained latent diffusion model, we introduce a lightweight procedure that requires no additional training: (i) a dimension-wise noise injection that targets latent channels most informative of instrument identity, and (ii) an early-step clamping mechanism that re-imposes the input’s melodic and rhythmic structure during reverse diffusion. The method operates directly on audio latents and is compatible with text/audio conditioning (e.g., CLAP). We discuss design choices, analyze trade-offs between timbral change and structural preservation, and show that simple inference-time controls can meaningfully steer pre-trained models for style-transfer use cases. Demo available at\footnote{Audio demos: \href{https://anon-audio-demo-25.github.io/audio_demo/}{anon-audio-demo-25.github.io/audio\_demo}}
\end{abstract}
\section{Introduction}
\label{sec:intro}
Timbre transfer alters the instrument identity of a \emph{source} recording—its timbre—while preserving the underlying musical \emph{content} (melody, harmony, rhythm). This capability is broadly useful in music production, e.g., it accelerates re-orchestration and arrangement cycles, expands creative control for sound design and remixing, or enables corrective editing when the performance is musically correct but the instrument color is not.

Recent advances in generative modeling—most notably diffusion probabilistic models—now produce high-quality, realistic audio and offer increasingly expressive conditioning interfaces \cite{Ho2020,Popov2021,Huang2022,Mittal2021,Huang2023,Popov2022,Choi2023}. Turning these general frameworks into a practical tool for instrument timbre transfer remains non-trivial. One prominent line trains dedicated systems for the task—often a separate model per target instrument or domain—which delivers strong quality but scales poorly and is narrowly tailored to timbre transfer \cite{LDBridge,WaveTransfer}. A second line repurposes powerful pretrained backbones by adding trainable post-hoc control modules to expose new control channels—such as independent melodic and timbral control, but at the cost of extra training and architectural changes \cite{hou2024stableaudiocontrol}. A third, training-free line steers sampling via guidance and related control signals, or performs per-example inference-time optimization of latent trajectories \cite{Ho2022cfg,Hertz2022,Zhang2023,Novack2024}. While effective, these strategies typically incur per-target optimization, added compute, or model modifications—motivating methods that recover fine-grained control directly at inference within a single pretrained model and no added computational cost.

We introduce a \emph{label-aware, optimization-free} method for \emph{inference-time} timbre transfer. It repurposes a pretrained latent diffusion model without retraining or per-example optimization, and runs at the computational cost of standard sampling. We instantiate the approach on \textsc{Diff-A-Riff}~\cite{Nistal2024,Nistal2024_2}, a latent diffusion system for musical accompaniment. The core challenge is that encoder latents entangle timbre (style) with structure (pitch, rhythm), and this entanglement varies by channel. We address this with a supervised mutual-information (MI) analysis against instrument labels to rank channels and derive masks. During sampling, we intervene in two ways: (i) inject noise only along timbre-aligned channels, and (ii) preserve—by clamping/restoring—structure-dominant channels across denoising steps. The result is a controlled shift in timbre toward the target while melody and rhythm remain intact.

Within this framework, our contributions are threefold: (i) a dimension-wise, MI-guided noise-injection mechanism that targets timbre-bearing channels; (ii) a clamping strategy that reinstates the original structural channels throughout the trajectory to safeguard timing and pitch; and (iii) an evaluation protocol that quantifies the trade-off between timbre change and content preservation. Although we focus on instrument timbre transfer as a well-labeled, practically compelling testbed, the same analysis-and-intervention recipe is general and could support other inference-time edits—e.g., denoising or enhancement conditioned on quality labels, performance style transfer driven by articulation/dynamics annotations, or broader label-based feature transfer wherever attributes can be measured or tagged.

\section{RELATED WORK}
Musical timbre/style transfer is often framed as domain adaptation: learn a mapping that changes instrument identity while preserving pitch and rhythm. Non-parallel cycle-consistency and related VAE–GAN variants have been adapted to musical audio \cite{huang2019timbretron, sammut2021ttvaegan, bitton2018move}. Diffusion is increasingly used inside the same mapping paradigm—training direct spectrogram/waveform translators or connecting source→target distributions in an audio latent \cite{comanducci2023difftransfer, WaveTransfer, LDBridge}. These models work well with a fixed, trained-on set of target domains, but typically rely on instrument labels or target-specific parameters (e.g., heads, embeddings, adapters). Moreover, achieving high fidelity and stable convergence generally requires substantial data, compute, and hyperparameter tuning.

A complementary thread exposes factorized controls so structure and timbre can be steered separately. DDSP autoencoders provide interpretable bottlenecks (F0, loudness, harmonic–noise) that support convincing swaps with little data \cite{engel2020ddsp}. Recent work attaches lightweight control branches to large generators \cite{wu2023musiccontrolnet, hou2024stableaudiocontrol, baker2025lilac}. These methods offer practical editing knobs while keeping a strong backbone mostly frozen, but they rely on reliable control extraction and paired control–audio during training, and bottleneck capacity/inductive bias can under-represent complex timbres and transients.

Finally, some approaches recover timbre control at inference by steering a pretrained generator’s sampling path—e.g., per-example optimization of the initial noise or time-varying inversion for stylization \cite{novack2024ditto, li2024musicti}. These methods avoid retraining and adapt to each input, but they can be tricky to tune; optimization-based variants add computation for every example, and without explicit constraints on structure, edits can drift and distort pitch or rhythm.

Our approach sits at this intersection: like domain-adaptation and factorized-control designs, it targets explicit timbre transfer; like inference-time steering, it requires no retraining—yet it also avoids per-example optimization by applying information-guided, channel-selective perturbations while clamping structure-dominated channels to preserve pitch and rhythm.

\section{Background}
Our work builds on Diff-A-Riff (DaR)~\cite{Nistal2024_2} (Sec.\ref{sec:dar2}), a generative model of musical stems in the latent space of Music2Latent2 (M2L2)\cite{Pasini2024} (Sec.\ref{sec:m2l}). We endow DaR with timbre transfer by performing a pre-hoc mutual-information analysis of M2L2 latents to disentangle structure from timbre-dominant dimensions, and intervene on the latter during the reverse diffusion trajectory. Although DaR natively employs EDM sampling, we also benchmark a deterministic DDIM baseline (Sec.\ref{sec:ddim}) for its convenient properties.

\vspace{0.2cm}
\inlineSubsection{Diff-A-Riff}{sec:dar2} \cite{Nistal2024_2} is a latent diffusion model designed for musical accompaniment generation, operating in the compressed latent space of the M2L2 encoder. 
Following the DiT (Diffusion Transformer) architecture, it supports a wide spectrum of generation modes, from unconditional to context-guided, using various inputs like CLAP embeddings~\cite{Elizalde2023}. A key advantage over many unconditional or symbolic models is its ability to offer fine-grained structural control while maintaining stylistic flexibility. Its training on diverse datasets enables robust one-shot generation, making it an excellent foundation for structure-preserving tasks like timbre transfer.

\vspace{0.2cm}
\inlineSubsection{Music2Latent2 (M2L2)}{sec:m2l} is a consistency autoencoder for audio reconstruction and conditioning. It encodes 48kHz stereo audio into a compact, low-frame-rate latent stream (11.7Hz, 64 channels). The encoder uses frequency-wise self-attention and per-band scaling to retain high-fidelity structural information despite the high compression. Meanwhile, its one-step consistency decoder enables fast waveform reconstruction with minimal artifacts. This excellent trade-off between compression and audio quality makes M2L2 highly suitable for resource-constrained generative pipelines.

\vspace{0.2cm}
\inlineSubsection{Denoising Diffusion Implicit Models (DDIM)}{sec:ddim}\cite{Song2020} generalize traditional diffusion models by allowing deterministic sampling trajectories, significantly reducing inference time without compromising generation quality. Further, DDIM enables inverting the mapping of an existing data point back to its latent representation by deterministically reversing the diffusion process. This property makes DDIM particularly suitable for tasks that require precise manipulation or preservation of structural attributes within the latent space, as in our proposed approach for timbre transfer.

\section{Methodology}

We propose a method for conveying timbre-transfer capabilities to DaR (see Sec.~\ref{sec:dar2}). In what follows, we describe the proposed method and the experimental setup, including Data, training details, and Evaluation method.
\begin{figure*}[t]
  \centering
  \includegraphics[width=0.9\linewidth]{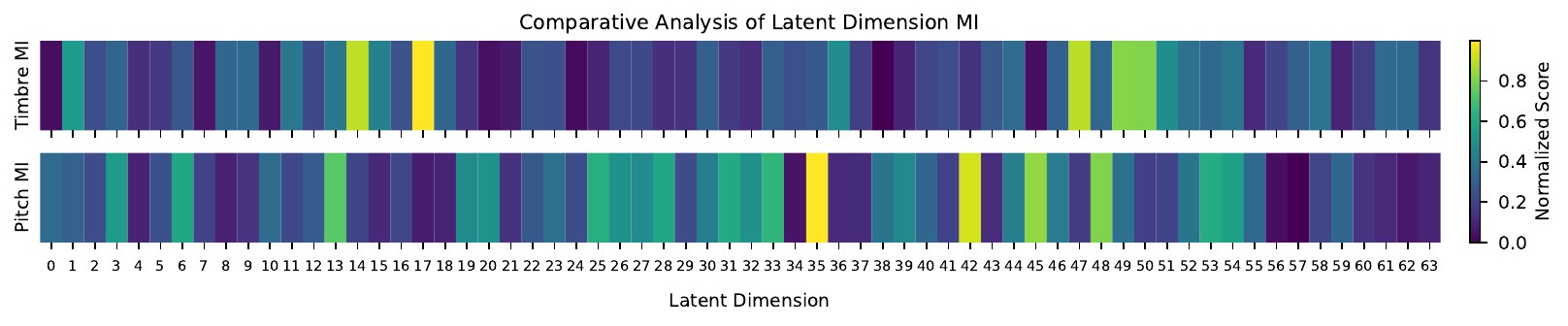}
  \vspace{-0.6cm}
  \caption{Per-channel mutual information between latent value and instrument class and pitch, and are normalised between [0, 1].}
  \label{fig:mi-heat}
\end{figure*}

\vspace{0.2cm}
\inlineSubsection{Dataset}{sec:dataset}
We use two corpora with complementary roles: (i) \textbf{NSynth} dataset \cite{Engel2017} for
\emph{analysis and tuning} (mutual information and baseline
noise calibration), and (ii) a \textbf{subset of the DaR testing
corpus} for \emph{timbre-transfer experiments and reporting}, by pairing tracks from different songs as context audio and reference audio.
Each evaluation clip is 10 seconds long and contains no more than 5 seconds of silence. We found that more than 7000 data pairs do
not yield a significant variance in our evaluation metrics. We
therefore adopt 7000 for parameter search.

\vspace{0.2cm}
\inlineSubsection{Baseline}{sec:baseline}
\vspace{0.2cm}

\inlineSubSubsection{Naive Method: Partial Noise Injection (PnI)}{sec:baseline}
A simple method for performing timbre transfer involves naively adding Gaussian noise to the \emph{source 
content} latents across all dimensions~\cite{Meng2022}:
\( \mathbf{z}_{t_f}^{partial} = z_{\text{in}} + \sigma_{t_f}\,\epsilon, \)
where $\epsilon\sim\mathcal{N}(0,I)$.
The target diffusion step $t_f$ is determined by finding the step closest to a user-defined noise fraction $f_{\text{par}} \in (0, 1]$:
\begin{equation*}
    t_f \;=\; \arg\min_{0\le t\le T}\, \bigl|\sigma_t - f_{\text{par}}\cdot\sigma_{\max}\bigr|,
\end{equation*}
where $\sigma_{\max} \equiv \sigma_T$. Here, $\sigma_{T}$ represents the maximum noise level of the EDM schedule (at step $T$), and \(f_{\text{par}}\) is determined by a user-defined
noise fraction parameter. We can use these expressions to look up the discrete timestep whose noise level \(\sigma_{t}\) lies closest to the desired fraction parameter \(f_{\text{par}}\) (e.g.\ \(f_{\text{par}}=0.5\) for ‘‘50 \% of the schedule’’). Because the EDM schedule \(\sigma_t\) is non-linear in \(t\) \cite{Karras2022}, this lookup yields a more faithful noise‐power match than a naive scaling \(z_{\text{in}}+f_{\text{par}}*\sigma_{\max}\).

\vspace{0.2cm}
\inlineSubSubsection{DDIM-based Partial Noise Injection (DDIM-PnI)}{sec:baselineddim}
We also explore a synthetic corruption with a deterministic DDIM inversion. We reuse the EDM grid $\{\sigma_t\}_{0..T}$ and map it to DDIM coefficients 
$\alpha_t=\tfrac{1}{1+\sigma_t^2}$,\; $1-\alpha_t=\tfrac{\sigma_t^2}{1+\sigma_t^2}$.
Starting from $x_{\mathrm{ctx}}$ at $t{=}0$, we run DDIM from noise level T and \emph{no guidance} (CFG$=0$) up to $t_f$ (chosen via $f_{\text{par}}$), yielding $z^{\mathrm{inv}}_{t_f}$.
Compared to the PnI ($z_{\text{in}}+\sigma_{t_f}\epsilon$), this variant only changes how the start state is obtained—via inversion rather than synthetic noise—while using the same schedule and step count.

\vspace{0.2cm}
\inlineSubSubsection{Partial Noise-Level Selection}{sec:noiseselect}
To set the PnI and DDIM-PnI noise fraction $f_{\text{par}}$, we adopt an Ambient Diffusion Omni–style probe \cite{Daras2025}.
We train a tiny timestep-conditioned binary classifier $C_\phi(z_t,t)$ to distinguish \emph{clean} NSynth latents from \emph{timbre-swapped} (ones swapping the timbre-related dimensions as identified via mutual information, see Sec~\ref{sec:dim_selection}).
For each candidate $f_{\text{par}}$, we corrupt latents to $t_f$ on the EDM grid and measure held-out accuracy. We choose the \emph{smallest} $f_{\text{par}}$ at which accuracy falls to chance ($\approx 50\%$), indicating instrument ambiguity from diffusion noise alone; this yields $f_{\text{par}}^\star \approx 0.5$ on our dev set, and suggested a starting point for later parameter search.

\vspace{0.2cm}
\inlineSubsection{Proposed Method MI-Guided Inpainting}{sec:proposed}. We propose an \emph{inference-time timbre editing} method that \emph{preserves musical structure} in a pre-trained latent diffusion model without added training or architectural changes. Given a \emph{context} audio \(x_{\mathrm{ctx}}\) (melody/rhythm to preserve) and a \emph{timbre target} specified by text or audio via a CLAP embedding,\footnote{\url{huggingface.co/lukewys/laion_clap}} we introduce two lightweight controls: (i) \emph{MI-guided, dimension-wise noise reinjection} that refreshes only the channels most informative of instrument identity (top-\(k\%\) by mutual information), and (ii) \emph{early-step clamping} that overwrites structure-dominant channels with the DDIM-inverted latent of \(x_{\mathrm{ctx}}\) during the high-noise regime of sampling (Sec.~\ref{sec:ddim}). Both operations incur negligible overhead and leave the sampler and weights unchanged.

\begin{figure}[t]
  \centering
  \includegraphics[width=0.8\linewidth]{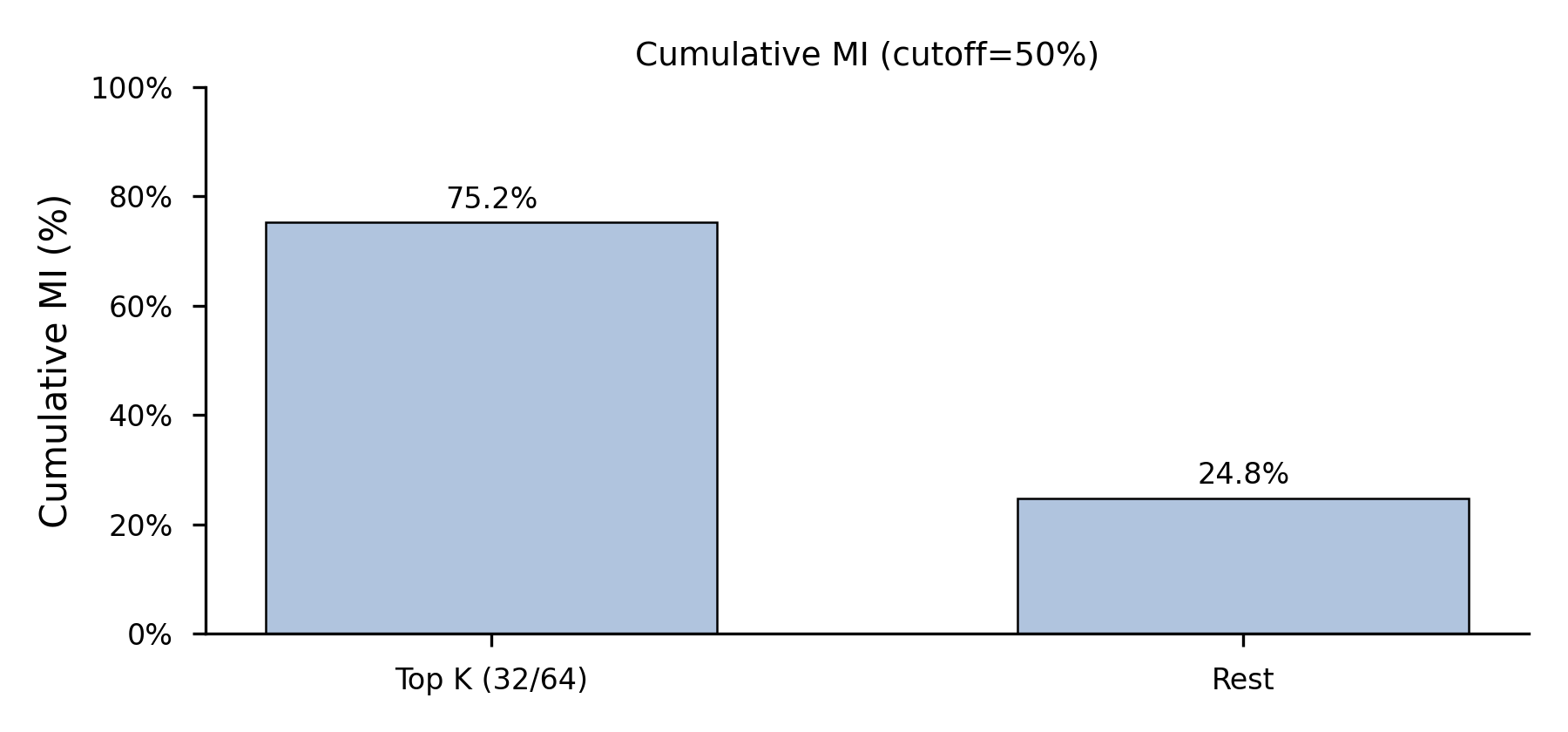}
  \vspace{-0.3cm}
  \caption{Cumulative MI for instrument identity at $k=0.5$: left, MI in the top 32 latent channels; right, residual MI in the remaining channels.}
  \label{fig:mi-bar}
\end{figure}

\vspace{0.2cm}
\inlineSubSubsection{Mutual Information (MI) Latent Analysis}{sec:dim_selection}
We perform MI analysis of non-silent M2L2 time-frames and instrument labels. Although M2L2 latents exhibit some degree of entanglement, Fig.~\ref{fig:mi-heat} demonstrates an emergent specialization where specific latent channels carry significantly higher information regarding instrument identity than others. This justifies selectively masking the top-k timbre-dominant dimensions to suppress timbre while preserving structural content. $\mathbf{M}_{\text{timbre}}$ or the top-k channels. The complementary mask, $\mathbf{M}_{\text{struct}} = 1-\mathbf{M}_{\text{timbre}}$ isolates the remaining dimensions. We treat the threshold  \(k\) as a hyperparameter to balance timbre-transfer strength with melodic preservation.
As shown in Fig.~\ref{fig:mi-bar}, the top-k channels capture most instrument-related MI, which justifies our grid-search choice $k{=}0.5$ (top 50\%) for $\mathbf{M}_{\text{timbre}}$.

\vspace{0.2cm}
\inlineSubSubsection{Dimension-Wise Noise Injection}{sec:dw_ni} :
Instead of drawing a fully random $z_{T}\sim\mathcal N(0,I)$, we construct a \textit{structured} start point that already contains the target melody and rhythm but allows the model to freely re‑synthesise timbre. Given the \emph{source content} waveform $x_{\mathrm{ctx}}$, we first obtain a latent $z^{\text{ctx}}\in \mathbb{R}^{1\times 64\times T'}$ by running \textsc{DDIM inversion} for $T$ steps with the same $\sigma_{t}$ grid that will be used during generation. Compared with the pure encoder output $E(x_{\mathrm{ctx}})$ used in the baseline, $z^{\text{ctx}}$ already has the same global variance $\mathrm{Var}[z]=\sigma_{T}^{2}$ as a standard EDM start state, eliminating the per‑dimension SNR mismatch with our unique latent initialization. Next, we perform a latent partition:
\begin{equation*}
  z_{t}^{\ast} \;=\;
  \underbrace{\sigma_{t}\,\epsilon\odot\mathbf{M}_{\text{timbre}}}_{%
      \text{pure noise on timbre dims}}
  \;+\;
  \underbrace{z^{ctx}_{\text{t}}
              \odot\mathbf{M}_{\text{struct}}}_{%
      \text{inverted latent on structure dims}}
  \label{eq:init-latent_dw}
\end{equation*}
At the start of denoising, the initial noise state becomes:
\(
  \mathbf{z}_{T}^{\text{tim}}
    = \sigma_{T}\,\boldsymbol{\epsilon}
       \,\odot\,\mathbf{M}_{\text{timbre}}; \quad
  \mathbf{z}_{T}^{\text{str}}
    = z_{T}^{\text{ctx}}
       \,\odot\,\bigl(1-\mathbf{M}_{\text{timbre}}\bigr); \quad
  \mathbf{z}_{T}^{\text{dim-wise}}
    = \mathbf{z}_{T}^{\text{tim}}
       + \mathbf{z}_{T}^{\text{str}},
\)
where \(z_{T}^{\text{ctx}}\) is the \(DDIM(x_{\text{ctx}})\) results in noisiest step T,
\(\boldsymbol{\epsilon}\sim\mathcal{N}(0,I)\) is freshly drawn Gaussian noise,  
\(\sigma_{T}\) is the EDM noise levels in the starting
step \(T\)  and
\(\mathbf{M}_{\text{timbre}}\in\{0,1\}^{64}\) is the binary mask that selects
the channels dominant in timbre \(k\). \(\mathbf{z}_{T}^{\text{tim}}\) is the pure-noise initialisation of the
timbre dimensions, \(\mathbf{z}_{T}^{\text{str}}\) is the
inverted latent of the structure dimensions, and their
sum \(\mathbf{z}_{T}^{\text{dim-wise}}\) is the complete latent fed to the reverse
diffusion process. We run the EDM solver for steps \(t=T,\dots,0\).
For the first part of the trajectory
(\(t\ge t_{c}\)), we clamp the structure dimensions: 
\begin{equation*}
  z_{t}
  \leftarrow
  z_{t-1}\odot\mathbf{M}_{\text{timbre}}
  \;+\;
  z_{t-1}^{ctx}\odot\mathbf{M}_{\text{struct}},
  \qquad
  t = T,\dots,t_{c}.
  \label{eq:clamp} 
\end{equation*}
The clamp-end step \(t_{c}\) is chosen by
\(\sigma_{t_{c}}\!\approx\!f_{\text{clamp}}\,\sigma_{T}\) with a grid search, keeping a certain amount of the clean signal, allows the model to follow the overall structure during denoising, mirroring the observation that early diffusion steps fix the global structure \cite{Ho2020}. Beyond \(t_{c}\) the sampler runs unconstrained, refining fine timbre details. 

\vspace{0.2cm}
\inlineSubsection{Inference}{sec:hp} We fix the sampler to \(N=30\) denoising steps and CFG strength to \(1.25\)~\cite{Nistal2024}. Inference parameters \(k\) (percentage of channels in the timbre mask \(\mathbf{M}_{\text{timbre}}\)), \(f_{\text{clamp}}\) (fraction of clamping step), and—baseline only—\(f_{\text{par}}\) (uniform noise fraction) are finetuned by means of a small grid search. We use CLAP timbre similarity and dynamic pitch distance(DPD) as performance indicators and keep the configuration that yields the best timbre similarity with minimal pitch drift. These hyperparameters are then fixed for all reported test results.

\vspace{0.2cm}
\inlineSubsection{Evaluation}{sec:evaluation}
We evaluate the proposed method with objective metrics against our training-free baselines (PnI, DDIM-PnI, DDIM-inversion) and a subjective listening study with trained reference system (WaveTransfer). We also include a small ablation to examine trade-offs of parameters.

\vspace{0.2cm}

\inlineSubSubsection{Metrics:}{sec:metrics} We evaluate timbre transfer quality using Fréchet Audio Distance (FAD) \cite{Kilgour2019} and CLAP similarity \cite{Elizalde2023}. Melodic preservation is assessed with two complementary metrics: Dynamic Pitch Distance (DPD) using the PESTO pitch-estimator \cite{Riou2023}, and note-onset F1. Note that while our qualitative demos include polyphonic examples to demonstrate robustness, objective metrics (DPD and Onset F1) were calculated strictly on monophonic segments of the test set, as PESTO is a monophonic estimator. We doubled the number of data pairs used for the parameter search to 14000 in this evaluation.

\vspace{0.2cm}
\inlineSubSubsection{Subjective test:}{} We ran a MUSHRA test assessing (i) timbre similarity, (ii) content preservation, and (iii) audio quality, comparing our MI-guided inpainting with WaveTransfer on demo excerpts and 10-s MIDI segments. We also collected MOS for realism on a 5-point scale (5=Realistic, 1=Unrealistic). Because the method exposes a timbre–structure trade-off, we set the noise-width $k$ and clamp fraction $f$ per example via a brief tuning pass to reach a balanced operating point, matching the intended user workflow.

\vspace{0.2cm}
\inlineSubSubsection{Ablation experiment:}{} We analyse the effect of the two hyperparameters $k$ (fraction of timbre-dominant channels) and $f_{\text{clamp}}$. A 3×2 grid ($k\in\{0.45,0.50,0.55\}$, $f_{\text{clamp}}\in\{0.40,0.45\}$) is evaluated on objective metrics, illustrating the trade-off between timbre similarity and melodic preservation.

\vspace{0.2cm}
\inlineSubSubsection{Baselines:}{sec:baselines}
We compare to three training-free baselines; here we only state the protocol (details cross-referenced).
\textbf{(1) DDIM-inversion.}
Deterministically invert the context from $t{=}0\!\to\!T$ (same EDM $\{\sigma_t\}$ mapped to DDIM, with CFG$=0$), then \emph{denoise to $t{=}0$ with the target CLAP}.
\textbf{(2) PnI.}
Add noise to the context at level $\sigma_{t_f}$ (with $t_f$ chosen by $f_{\text{par}}$ on the same EDM grid), then \emph{EDM denoise from $t_f\!\to\!0$ with the target CLAP}.
\textbf{(3) DDIM-PnI}
Same as DDIM-inversion, but invert only to an \emph{intermediate} $t_f$ (not to $T$), then \emph{EDM denoise from $t_f\!\to\!0$ with the target CLAP}.
All runs use the same $\{\sigma_t\}$, $N{=}30$ steps, CFG$=1.25$ at generation, identical CLAPs (only the target is swapped), loudness norm, and the same context–reference pairs; no per-example tuning. For subjective tests we also report \textbf{WaveTransfer-BDDM19}. We did not include the BDDM19 model in the objective evaluations, as it was trained on a limited number of instruments, incompatible with our evaluation set.

\section{Results and discussion }
\begin{table}[h]
  \centering
  \footnotesize
  \caption{Objective results. \underline{Underlined} = best across all methods; \textbf{bold} = best within our method. (↓ lower is better, ↑ higher is better). }
  \label{tab:objective}
  \setlength{\tabcolsep}{2pt}

  \begin{tabular}{
      l
      S[table-format=1.2] 
      S[table-format=1.2] 
      S[table-format=1.3] 
      S[table-format=3.2] 
      S[table-format=1.3] 
      S[table-format=1.2] 
    }
    \toprule
    & \multicolumn{2}{c}{\textsf{Params}} & \multicolumn{4}{c}{\textsf{Metrics}}\\
    \cmidrule(lr){2-3}\cmidrule(lr){4-7}
    Method & {$k$} & {$f$} &
    {\textsf{FAD}\,↓} &
    {\textsf{DPD}\,(\textcent)\,↓} &
    {\textsf{CLAP}\,↑} &
    {\textsf{F1 Onset}\,↑} \\
    \midrule
    PnI            & \multicolumn{2}{c}{--} & 3.74 & 110.79 & 0.63 & 0.37 \\
    DDIM-PnI       & \multicolumn{2}{c}{--} & 1.48 & \multicolumn{1}{c}{\underline{29.71}} & 0.52 & \multicolumn{1}{c}{\underline{0.78}} \\
    DDIM-inversion & \multicolumn{2}{c}{--} & \multicolumn{1}{c}{\underline{1.333}} & 196.93 & \multicolumn{1}{c}{\underline{0.76}} & 0.14 \\
    \midrule
    Setting 1 & 0.55 & 0.40 & \bfseries1.65 & 108.70 & \bfseries0.61 & 0.49 \\
    Setting 2 & 0.55 & 0.45 & 1.72 & 104.89 & 0.60 & 0.50 \\
    Setting 3 & 0.50 & 0.40 & 1.73 & 103.23 & 0.60 & 0.58 \\
    Setting 4 & 0.50 & 0.45 & 1.75 &  97.96 & 0.59 & 0.59 \\
    Setting 5 & 0.45 & 0.40 & 1.74 &  83.64 & 0.57 & 0.63 \\
    Setting 6 & 0.45 & 0.45 & 1.79 & \bfseries 78.82 & 0.56 & \bfseries 0.64 \\
    \bottomrule
    (high-$k$, long-$f$) & 0.6 & 0.5 & 1.56 &  133.20 & 0.65 & 0.48 \\
    (high-$k$, short-$f$) & 0.6 & 0.3 & 1.50 &  159.75 & 0.69 & 0.43 \\
    (low-$k$, long-$f$) & 0.4 & 0.5 & 1.81 & 56.04 & 0.54 & 0.71 \\
    (low-$k$, short-$f$) & 0.4 & 0.3 & 1.71 & 88.95 & 0.57 & 0.62 \\
    \bottomrule
  \end{tabular}
\end{table}

\vspace{0.2cm}
\inlineSubsection{Objective Evaluation}. EDM partial denoising (PnI) is the most straightforward way to perform timbre transfer with the backbone model, but it harms content: DPD is high and F1-Onset is low. Although CLAP is strong—i.e., timbre transfer occurs—audio quality degrades, as reflected by high FAD. Replacing EDM with DDIM while keeping the same partial noise-injection recipe (reversing the source latent partway through the schedule) markedly improves content preservation (lower DPD, higher F1-Onset) and overall quality (lower FAD), albeit with weaker timbre transfer (lower CLAP). Extending the reverse process to the full schedule (DDIM-inversion) recovers timbre transfer (higher CLAP) and further improves FAD, but again challenges content preservation. These baselines reveal an entanglement in Diff-A-Riff between CLAP-guided timbre conditioning and the generated structure. Our approach—MI-guided dimension-wise noise injection with early clamping—shifts this trade-off toward stronger structural preservation while keeping timbre transfer competitive. Intuitively, increasing the number of perturbed channels 
$k$ injects stochasticity into timbre-dominant dimensions and boosts stylistic expressiveness; because MI is not perfectly disentangled, very large 
$k$ also perturbs mixed-role channels and invites pitch or onset drift. Early clamping anchors melody and event timing at the stage where global structure is decided, while overly long clamping can constrain later fine-grained timbral updates, so $k$ and clamp length $f$ counterbalance one another and create a knee point where a small concession in timbre yields a disproportionate gain in structure. The table reflects these mechanisms: as $k$ decreases and 
$f$ increases from Setting 1 to 6, structure improves consistently—DPD drops and F1-Onset rises with only modest CLAP reduction and a small FAD increase. Mid-range choices (Settings 3–4) provide a robust operating point that balances timbre transfer and structural fidelity without incurring large FAD penalties; CLAP tends to rise with $k$ and then saturate, while DPD/F1 track the strength of structural anchoring, and FAD—being distributional rather than timbre-specific—need not mirror perceived timbre changes. In short, dimension-wise noise with early clamping enables inference-time timbre transfer on a pre-trained diffusion model and improves the CLAP–DPD balance over uniform SDEdit and DDIM partial-noise baselines; the best trade-off occurs at mid-range $k$ and clamp duration, consistent with the view that early reverse-diffusion steps set structure while later steps realize timbre. 

\begin{table}[t]
  \centering
  \footnotesize
  \caption{Subjective listening contrasts (Ours vs.\ BDDM19) from a mixed-effects model.
  $\Delta$ is the fixed-effect coefficient (Ours $-$ BDDM19); negative values favor BDDM19 for the named attribute, positive values favor Ours. $p_{\text{holm}}$ are Holm-corrected.}
  \label{tab:subjective-contrasts}
  \setlength{\tabcolsep}{6pt}
  \begin{tabular}{
    l
    S[table-format=+1.3]  
    S[table-format=+2.2]  
    S[table-format=<1.1e-2] 
    c                     
  }
    \toprule
    \textsf{Block} & {$\Delta$ ($\beta$)} & {$z$} & {$p_{\text{holm}}$} & $n_{\text{list}}\!\times\!n_{\text{items}}$ \\
    \midrule
    Timbre  & -0.395 & -4.37 & 1.25e-5  & $24\times 5$ \\
    Content & -0.035 & -0.57 & 0.568    & $24\times 5$ \\
    Quality & +0.627 & +17.31 & <1e-6   & $24\times 5$ \\
    \bottomrule
  \end{tabular}
\end{table}

\begin{table}[t]
  \centering
  \footnotesize
  \caption{Mean Opinion Score (MOS) for realism on a 1--5 Likert scale
  (\textbf{maximum level = 5 = most realistic})}
  \label{tab:mos_realism_rev}
  \begin{tabular}{lccc}
    \toprule
    System & MOS $\uparrow$ & 95\% CL (1--5) & $n$ \\
    \midrule
    Ours          & 3.52 & {[$\text{3.37}$, $\text{3.68}$]} & 115 \\
    BDDM19  & 2.10 & {[$\text{1.94}$, $\text{2.25}$]}     & 115 \\
    MIDI-Reference  & 4.04 & {[$\text{3.85}$, $\text{4.23}$]}     & 115 \\
    \bottomrule
  \end{tabular}
\end{table}

\vspace{0.2cm}
\inlineSubsection{Subjective Evaluation}.
We ran a MUSHRA-style test with 29 listeners on 60 excerpts. For each contrast, the null hypothesis was \(H_{0}:\,\beta_{\text{ours}}=0\) (no difference in mean within-listener \(z\)-score vs.\ BDDM19), tested two-sided with Wald \(z\)-tests and Holm correction within block. Compared to the BDDM19, our proposed method achieved \emph{significantly higher} perceived audio quality (p-value $<0.001$) and showed \emph{no significant} difference in content preservation (p-value $>0.5$). Conversely, BDDM19 was rated \emph{significantly higher} for timbre similarity (p-value $<0.001$), indicating a trade-off consistent with the objective results. On the other hand, the MOS for perceptual realism indicates that our proposed method is most often rated \emph{Slightly Realistic} and BDDM19 is largely perceived as \emph{Slightly unrealistic} or \emph{Unrealistic}. This indicates that our method produces generally realistic renderings and is perceived as substantially more natural than WaveTransfer.

\vspace{-0.1cm}
\section{Conclusion}
We introduced a training-free timbre transfer method that steers a pretrained latent diffusion model using MI-guided, dimension-wise noise injection and early-step clamping. By perturbing timbre-dominant channels while reinstating structure-dominant ones early in sampling, our approach preserves melody/rhythm (lower DPD, higher onset F1) with competitive timbre similarity to baselines. The procedure is lightweight, architecture-agnostic, and compatible with text/audio conditioning. Remaining limitations include imperfect channel disentanglement and hand-tuned clamping; future work will explore adaptive masks or learned clamping schedules.

\begingroup
\small
\bibliographystyle{IEEEbib}
\bibliography{refs}
\endgroup

\end{document}